\documentclass[conference]{IEEEtran}
\IEEEoverridecommandlockouts
\usepackage{cite}
\usepackage{amsmath,amssymb,amsfonts}
\usepackage{algorithmic}
\usepackage{graphicx}     
\usepackage{textcomp}
\usepackage{xcolor}
\usepackage[hidelinks]{hyperref}
\usepackage{subfig}
\usepackage{booktabs, multirow} 
\usepackage{array, tabularx}
\usepackage{flushend}

\DeclareMathAlphabet\mathcal{OMS}{cmsy}{m}{n}
\DeclareMathAlphabet\mathbfcal{OMS}{cmsy}{b}{n}

\def\BibTeX{{\rm B\kern-.05em{\sc i\kern-.025em b}\kern-.08em
    T\kern-.1667em\lower.7ex\hbox{E}\kern-.125emX}}

\usepackage{fancyhdr}
\fancypagestyle{firstpage}{
  \fancyhf{}
  \fancyhead[C]{\small To be presented at the \textit{IEEE International Workshop on Multimedia Signal Processing (MMSP)}, Poitiers, France, Sept. 27 -- 29, 2023.}
}

\begin{document}

\title{Metaverse: A Young Gamer's Perspective\\
}

\author{
\IEEEauthorblockN{Ivan V. Baji\'{c}}
\IEEEauthorblockA{\textit{Simon Fraser University} \\
Burnaby, BC, Canada \\
}
\and
\IEEEauthorblockN{Teo Saeedi-Baji\'{c} and Kai Saeedi-Baji\'{c}}
\IEEEauthorblockA{\textit{\'{E}cole Pauline Johnson} \\
West Vancouver, BC, Canada \\
}
}

\maketitle

\begin{abstract}
When developing technologies for the Metaverse, it is important to understand the needs and requirements of end users. Relatively little is known about the specific perspectives on the use of the Metaverse by the youngest audience: children ten and under. This paper explores the Metaverse from the perspective of a young gamer. It examines their understanding of the Metaverse in relation to the physical world and other technologies they may be familiar with, looks at some of their expectations of the Metaverse, and then relates these to the specific multimedia signal processing (MMSP) research challenges. The perspectives presented in the paper may be useful for planning more detailed subjective experiments involving young gamers, as well as informing the research on MMSP technologies targeted at these users.  
\end{abstract}


\thispagestyle{firstpage}

\section{Introduction}
\label{sec:introduction}
Metaverse is expected to be the next iteration of the Internet. Building on advances in communications and networking, computer graphics, multimedia signal processing (MMSP), and other technologies, the Metaverse\footnote{We capitalize Metaverse when used as the noun (e.g., the Metaverse) and show it lowercase when used as an adjective (e.g., metaverse gaming).} will become the place for interaction, sharing, gaming, and a variety of services operating in virtual worlds. 
Several recent reviews of the Metaverse have appeared in the literature. In~\cite{2022_Metaverse_survey_arxiv, 2023_Metaverse_survey_CST,2022_Metaverse_security_privacy}, the authors provide a general overview of metaverse technologies with an  emphasis on security and privacy. Human-centric perspectives on  metaverse technologies are explored in~\cite{2023_human-centric_metaverse}, while~\cite{2023_Metaverse_healthcare,2022_Metaverse_medical} review  metaverse technologies related to healthcare, and~\cite{2022_Metaverse_education, 2022_Metaverse_learning} explore the role of the Metaverse in education.

The focus of this paper is on the perspectives of young (10-year-old and younger) persons on metaverse gaming.\footnote{User experience in  games has been studied in the past but, to our knowledge, most studies have targeted older users, e.g., 11+ years of age~\cite{UX_Access2018}.} This is important for several reasons. First, young gamers are among the first adopters of the Metaverse. Although the financial volume of this segment of users may be questioned because they do not control their own finances, the commercial success of gaming, in general, means that there are significant commercial opportunities in this market. Second, the perspectives of young people often differ from those of adult users, and this is generally underappreciated in the technical community. Finally, running user studies involving young users is more complicated than with adult users due to more stringent regulatory requirements, the need to involve parents/guardians, and their generally lower attention span that could impact the duration and logistics of such studies. This means that user studies that involve young users should be planned even more carefully than those involving adults, and the selection of hypotheses to be tested should be done judiciously. Our hope is that the present manuscript will aid in selecting such hypotheses for further studies. 

We will use Roblox,\footnote{\url{www.roblox.com}} the most popular metaverse gaming platform, for concrete illustrations. 
Roblox is a multi-player gaming platform where developers can create their own games, and users can select from a variety of developed games to play on their own or with their friends. Most Roblox games involve users' avatars, which themselves represent an important focal point for young gamers.    

This paper was written based on discussions with the last two authors, who are avid young Roblox gamers and who contributed some of their Roblox creations to the manuscript. While the discussions were mostly nontechnical, an attempt was made to map their perspectives and expectations into technical terms and relate them to relevant MMSP technologies. No new methodologies or algorithms are presented in this paper; instead, the goal is to illuminate how young gamers see metaverse gaming, what their expectations are of these technologies, and how these expectations can be addressed through MMSP research. Section~\ref{sec:prelim} presents some of the preliminaries needed for a subsequent discussion. Section~\ref{sec:interaction} discusses interaction with the Metaverse, which is the centerpiece of young gamers' view of the Metaverse. Section~\ref{sec:safety}  discusses health, safety, security, and privacy, 
followed by conclusions in Section~\ref{sec:conclusions}.

\section{Preliminaries}
\label{sec:prelim}
\begin{figure}[t]
    \centering
    \includegraphics[width=\columnwidth]{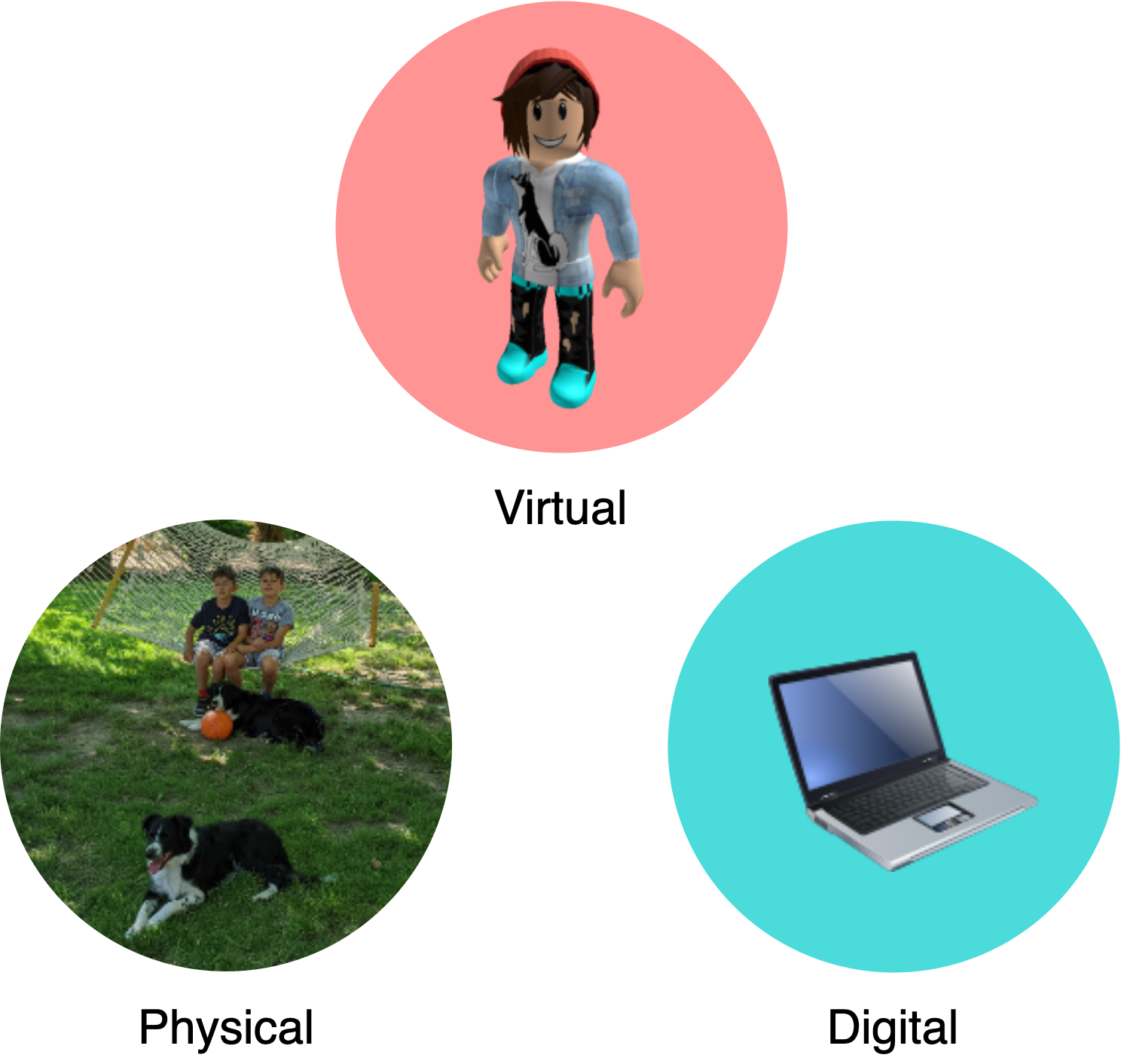}
    \caption{The three worlds discussed in the paper.}
    \label{fig:3_worlds}
\end{figure}

Through discussions with the last two authors, it became clear that their understanding of the Metaverse is shaped by their prior experience with two other worlds: \emph{physical} and \emph{digital}. The physical world is where their physical body resides, while the digital world is the technology they got used to prior to taking up metaverse gaming and includes digital images and video, simpler forms of gaming, as well as simple software. The Metaverse appears as the third world, which we will refer to as \emph{virtual}. Fig.~\ref{fig:3_worlds} illustrates the three worlds.


\begin{figure}[t]
    \centering
    \includegraphics[width=\columnwidth]{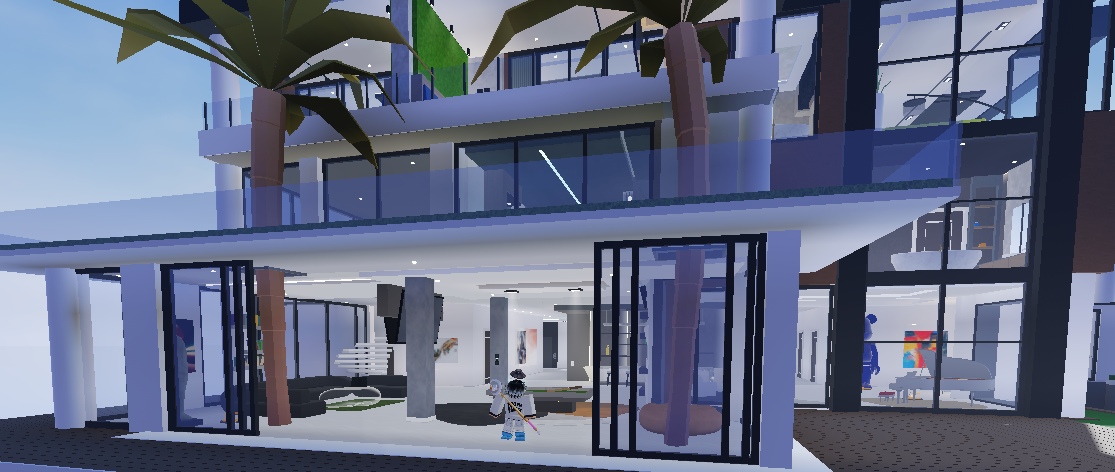}
    \includegraphics[width=\columnwidth]{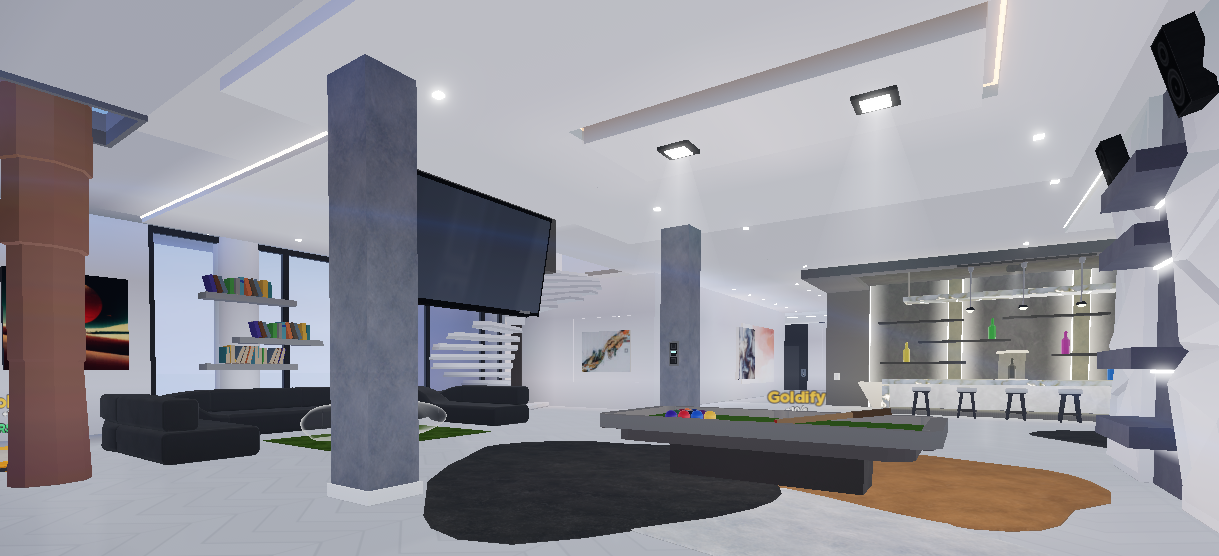}
    \caption{Top: penthouse; Bottom: living room.}
    \label{fig:penthouse}
\end{figure}

Young people have limited control over their life in the physical world, so the virtual world provides an environment in which they can exercise greater control and pursue their interests. Once they learn that many aspects of the virtual world are not limited by the same constraints as the physical world, it opens up many possibilities to exercise their imagination and creativity. One of the young gamers in this study took an interest in designing avatar clothing (Fig.~\ref{fig:3_worlds} top), because the opportunities to choose his clothing in the physical world were limited; the other was unhappy with the layout of his dwellings so he took up designing houses in the virtual world (Fig.~\ref{fig:penthouse}). We will use avatar clothing design to illustrate many of the MMSP research challenges related to the Metaverse. Table~\ref{tab:acronyms} lists the acronyms used in the paper.  

\begin{table}[t]
    \centering
    \caption{Acronyms used in the paper.}
    \label{tab:acronyms}
    \begin{tabularx}{0.36\textwidth}{>{\raggedright\arraybackslash}m{1.5cm} | >{\raggedright\arraybackslash}m{4.5cm}  }
    \toprule
       \textbf{Acronym}  & \textbf{Meaning} \\ \midrule
       MMSP & MultiMedia Signal Processing \\ \hline
       PNG & Portable Network Graphics \\ \hline
       JPEG & Joint Photographic Experts Group \\ \hline
       UGC & User-Generated Content \\ \hline
       QoE & Quality of Experience \\ \hline
       RGB+D & Red, Green, Blue plus Depth \\ \hline
       LiDAR & Light Detection and Ranging \\ \hline
       PC & Point Cloud \\ \hline
       VAC & Vergence-Accommodation Conflict\\ \hline
       SSO & Single Sign-On \\ \hline
       MFA & Multi-Factor Authentication \\
    \bottomrule
    \end{tabularx}
\end{table}

\section{Interactions among the worlds}
\label{sec:interaction}

The current generation of young gamers has grown up surrounded by touchscreens; for them, interaction with the digital world is fairly seamless, and the expectation is for such seamlessness to carry over to the virtual world. Just as moving an icon across a screen is effortless, it should be easy to move objects into and out of the virtual world. However, for now, this is far from easy; some of the key research problems stem from such challenges. 

\subsection{Interaction between digital and virtual worlds}

\begin{figure}[t]
    \centering
    \includegraphics[width=\columnwidth]{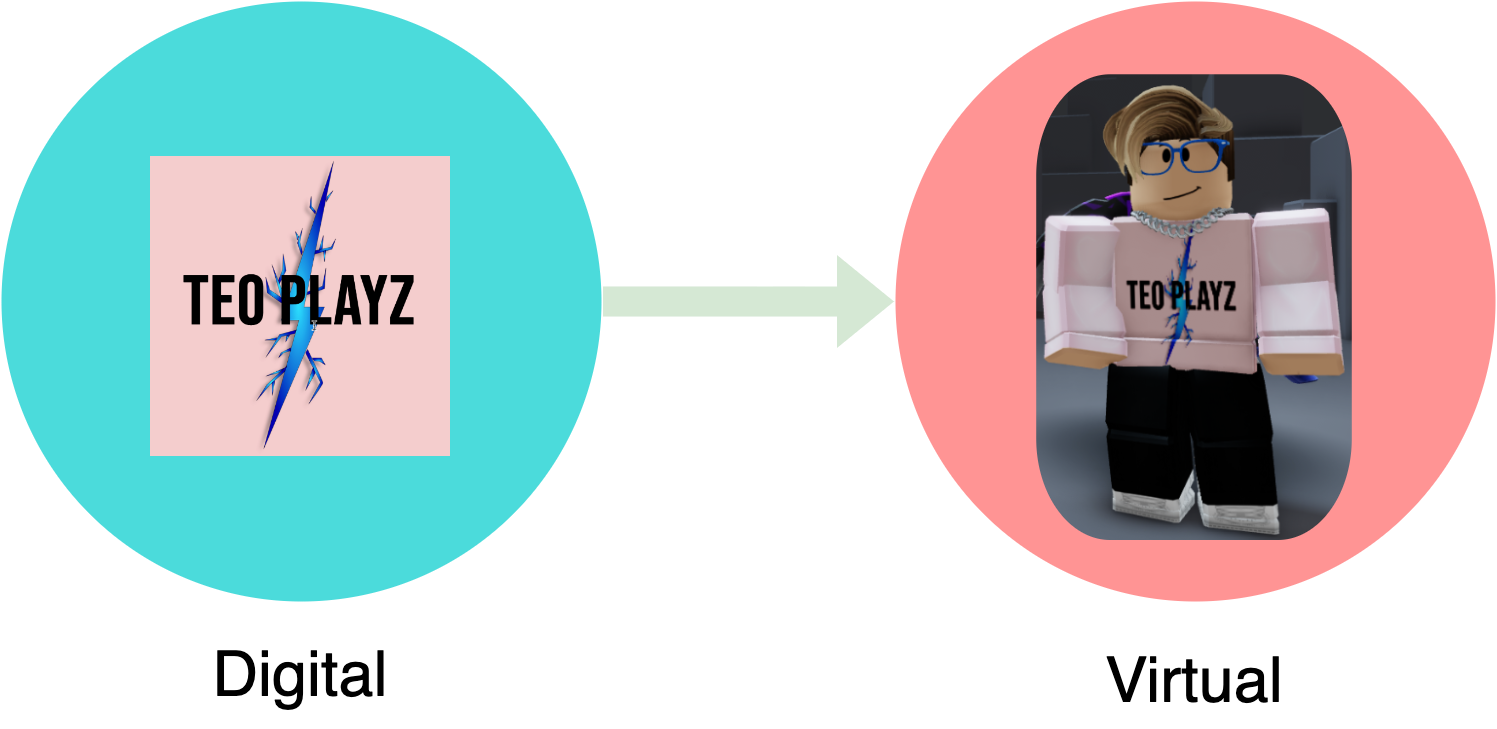}
    \caption{Transferring objects (e.g., a logo) from the digital world to the virtual world is relatively easy, though not seamless.} 
    \label{fig:digital_to_virtual}
\end{figure}

We use avatar clothing design as a motivating application to illustrate some of the challenges. Roblox offers many pre-made clothing items for avatars, but personalizing avatar clothing usually means having to design patterns or logos outside of Roblox. This can be done in a variety of image-editing software in the digital world. Fig.~\ref{fig:digital_to_virtual} shows an example of a logo designed using a simple image editor transferred to the virtual world. Roblox can import a number of image formats (PNG, JPEG, etc.) and transferring from the digital to the virtual world is fairly easy. The key technical question here is related to how image compression artifacts (for compressed image/video formats) would affect the quality of rendering the corresponding image/video in the virtual world. 
These issues have recently been studied, both in terms of conventional coding~\cite{HEVC_VR_videos} and foveated coding~\cite{foveated_compression_VR}.

There is also a need for moving data the other way: from the virtual to the digital world. In particular, user-generated content (UGC) videos of games have been popular on YouTube and other video sharing platforms, and the quality assessment of these videos has been the subject of recent studies~\cite{gaming_UGC_WACV,quality_gaming_TIP}. UGC videos often have both synthetic (the game itself) and natural content (e.g., the player's headshot), which highlights the need for codecs that are able to handle both synthetic and natural components. Screen content coding has been addressed in the latest video coding standards~\cite{HEVC-SCC,peng2016overview,bross2021overview,xu2021overview} as well as recent learning-based pre/post-processing~\cite{SR_game_video_ICASSP2023} and compression~\cite{multitask_SCC_ISCAS2023}. When encoding game video for video sharing platforms, one is likely to encounter content that is already encoded in the virtual world. Therefore, an efficient solution would be to transcode such content into a conventional video format. However, there appears to be little work on transcoding for such scenarios.


In terms of digital-virtual world interaction, there is also the question of playing the game on different devices. To an outsider, it might seem like a good idea to strive to make the playing experience on different devices (e.g., computer and tablet) as similar as possible. However, perhaps surprisingly, both gamers involved in this study indicated that they \emph{liked the fact that different devices provided different experiences} and that they would choose a device based on the type of experience they preferred on a given occasion. This point should be further examined in a larger user study.  

\subsection{Interaction between physical and virtual worlds}

As the virtual world resembles and often tries to mimic the physical world, it seems natural to want to transfer objects between the two worlds. Indeed, the increasing sophistication of avatars in metaverse games is an attempt to more closely mimic physical bodies. However, object transfer between the physical and the virtual worlds is the source of some of the greatest challenges related to metaverse technologies. 

In the context of our clothing example, Roblox offers advanced options for virtual clothing design through its Roblox Studio tool. These can incorporate designs created in other software, as described in the previous section. However, once a piece of clothing is designed in the virtual world, there is no easy way to transfer it to the physical world. The necessary steps would involve choosing the right size and shape (virtual try-on) and then choosing and cutting the material, sowing, and printing in the physical world. In the case of a more general object transfer from the virtual to the physical world, related steps would include sizing, material selection, and manufacturing (e.g., 3D printing). There has been quite a bit of research on virtual try-on in recent years~\cite{VITION_CVPR2018,OccluMix_TMM2023,AI_Fashion_SPM2023}. The same goes for 3D printing and advanced manufacturing, although these areas are not usually considered part of MMSP research.   

\begin{figure}[t]
    \centering
    \includegraphics[width=\columnwidth]{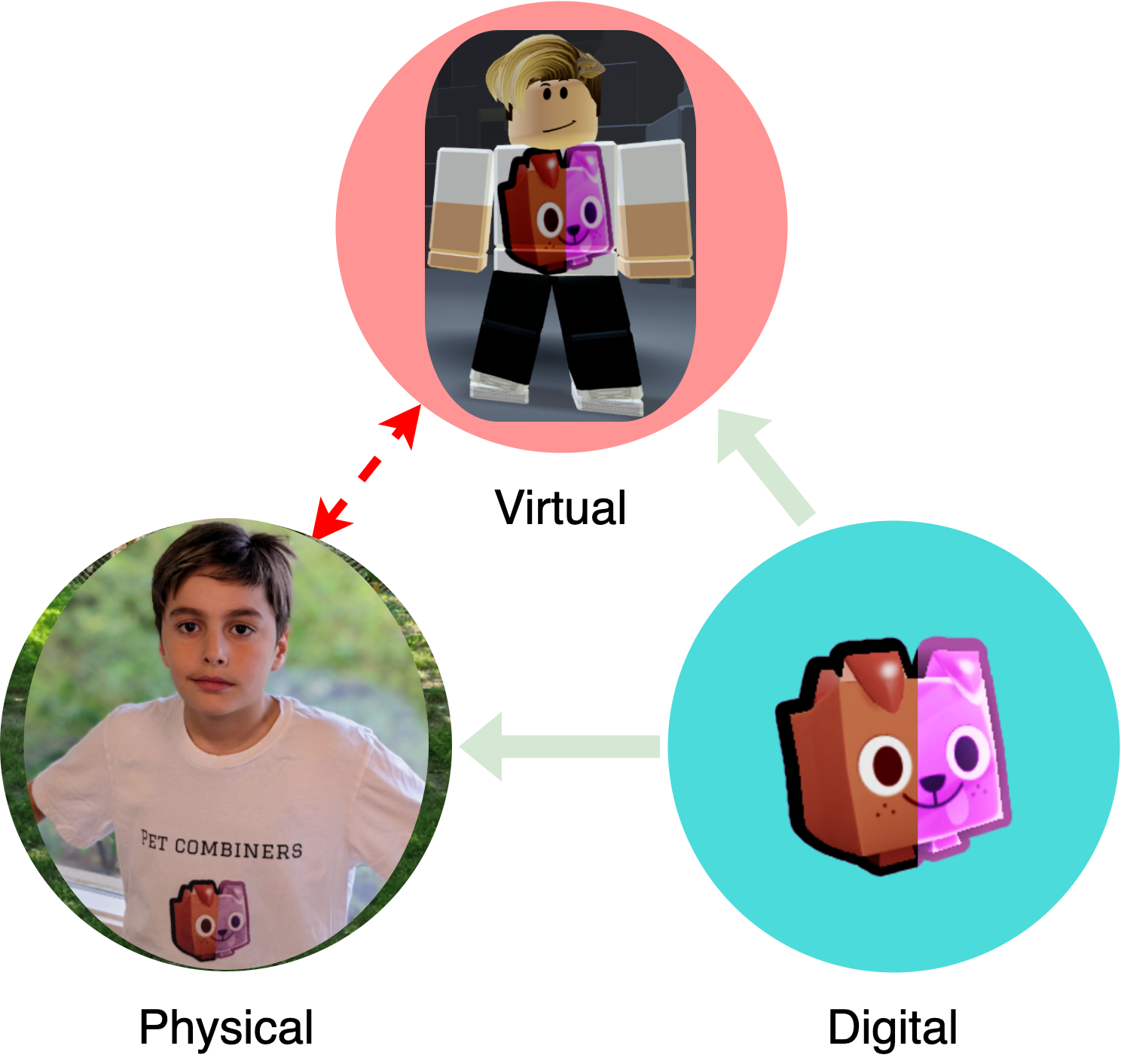}
    \caption{Transferring items between virtual and physical worlds (red dashed line) is challenging. The workaround is to transfer from digital to both virtual and physical.} 
    \label{fig:digital_to_virtual_and_physical}
\end{figure}

The workaround to the challenges of moving objects from the virtual to the physical world is to design them in the digital world and then transfer them to both the virtual and the physical world. An example with the t-shirt logo is shown in Fig.~\ref{fig:digital_to_virtual_and_physical}, and utilizes a t-shirt printing service (for digital $\to$ physical transfer) and Roblox Studio (for digital $\to$ virtual transfer). This workaround is, of course, cumbersome and making this process easier is among the greatest expectations of the young gamers involved in this study. 

The transfer from the physical to the virtual world has been more researched, although it remains challenging. The main research keywords in this area would be 3D capture, reconstruction, and compression of 3D models. Great strides have been made in the last few decades in related technologies. In terms of sensing, Red-Green-Blue plus Depth (RGB+D), multiview imaging, and Light Detection And Ranging (LiDAR) have all been popular topics in the MMSP community. RGB+D sensing, for example, using a Kinect-like sensor, offers a cost-effective way of obtaining 3D scans~\cite{RGBD_CVPR2015,RGBD_ICRA2019}. However, this approach generally offers lower accuracy compared to the alternatives. Multiview 3D reconstruction~\cite{Hartley2004}, LiDAR and industrial laser-based 3D scanners are more accurate but also more expensive options. For a young gamer who wants to transfer a 3D object from the physical to the virtual world, an ideal solution would be to use the same device they play the game on (e.g., a smartphone or tablet) and a process that takes only a few clicks. Technology is not there yet, although progress is being made.     

Point clouds (PCs) are a popular data format for representing 3D objects. PC processing, compression, and quality assessment are among the most active metaverse-related research thrusts in the MMSP community. In terms of PC processing, sampling~\cite{PC_sampling_TPAMI2023} and resampling~\cite{PC_resampling_MMSP2021}, denoising~\cite{PC_denoising_TIP2020}, registration~\cite{PC_registration_TMM2023}, super-resolution~\cite{PCSR_TIP2022},  and analysis~\cite{3Dseg_ICME2017,GeoClass_TMM2022} have attracted considerable research interest. For PC compression, video-based~\cite{MPEG-PCC_SPM2019} and geometry-based approaches~\cite{MPEG-PCC_JETCAS2019,PCGC_MMSP2022} have been developed. PC quality assessment addresses both geometry and color~\cite{PCQE_QoMEX2019,PointXR_QoMEX2020,PCQA_MMSP2021}. Both full-reference~\cite{PCQM_QoMEX2020} and no-reference~\cite{PCQA_MMSP2022} methods have been proposed. 

Finally, when discussing interactions between the physical and virtual worlds, it is worth mentioning latency and game responsiveness. Multiplayer games may involve hundreds of players distributed across the world, creating enormous challenges in terms of the need to provide low-latency interaction. Although dealing with latency is usually the purview of multimedia communications, the MMSP community has addressed related challenges in game video compression~\cite{Foveated_game_MMSP2017,Game_video_MMSP2021} and quality assessment~\cite{QoE_game_MMSP2014,GAMIVAL_SPL2023}.

\section{Health, safety, security, and privacy}
\label{sec:safety}

Young gamers have limited understanding of the health, safety, security, and privacy aspects of metaverse gaming. Hence, this section is mostly written from the parents'/guardians' perspective.  

\subsection{Health}
When it comes to health, parents/guardians of young gamers are usually concerned about screen time and its effect on focus and sleep~\cite{mobile_sleep_2020}. Another area of concern is the blue light emitted from screens and its long-term effect on vision~\cite{BLH_JLR2023}. Although there does not appear to be much MMSP research on these topics, one health-related area where the MMSP community has made notable contributions is the study of the vergence-accommodation conflict (VAC)~\cite{VAC_JoV_2008} in stereoscopic 3D  displays~\cite{VAC_QoMEX2011,VAC_TIP2016}. This phenomenon, which exists in 3D display systems that project objects at different depths onto the same focal plane, has been shown to cause visual fatigue and discomfort. The proposed solutions include multifocal~\cite{multifocal_ICME2016,multifocal_ICMEW2020} and time division multiplexed~\cite{CTDM_TIP2023} light field displays. Related quality assessment of light field content has also been a subject of intense research~\cite{LF_perceptual_QoMEX2017,LFQA_TIP2020}.

\subsection{Safety}
Safety issues for young gamers in the Metaverse center on two main concerns: (1) Who are they communicating with? and (2) Are they exposed to harmful / inappropriate content? MMSP research has addressed both concerns. In terms of identity verification, biometrics are a key technology, and the MMSP community has a long history of research in this area. Among the many approaches investigated for person identification are those based on face images~\cite{face_rec_MMSP2007}, head and mouth dynamics~\cite{head_mouth_MMSP2006}, gait~\cite{gait_MMSP2004}, audio-visual cues~\cite{AV_MMSP1998,AV_MMSP2009}, apparel~\cite{ApparelNet_MMSP2021}. The other issue - detection of inappropriate or harmful content - has been studied in both conventional images~\cite{adult_ICME2009,adult_ICME2011} and video~\cite{harmful_MMSP2014,adult_neurocomputing2017}. However, there does not yet appear to be any related published work in the specific context of the Metaverse.

\subsection{Security}

Some young gamers have heard of \emph{hackers} and \emph{hacking}. Although they lack technical understanding of these terms, they understand that someone else may take over their account, thereby also taking away their metaverse creations. Also, if their friend's account gets compromised, they may end up communicating with an unknown person thinking that they are their friend. So, in this general sense, they appreciate that security in the Metaverse is important and is not guaranteed. 

Account security has been evolving in response to various cyber threats. The current practice of single sign-on (SSO) federated identity management and multi-factor authentication (MFA) is likely to need upgrades in the future as the threats themselves evolve. Biometrics (face and fingerprints) have also been incorporated into account security, especially on mobile devices. This is an area where the MMSP community has contributed important research, as discussed above.

\begin{figure*}
\centering
\includegraphics[width=\textwidth]{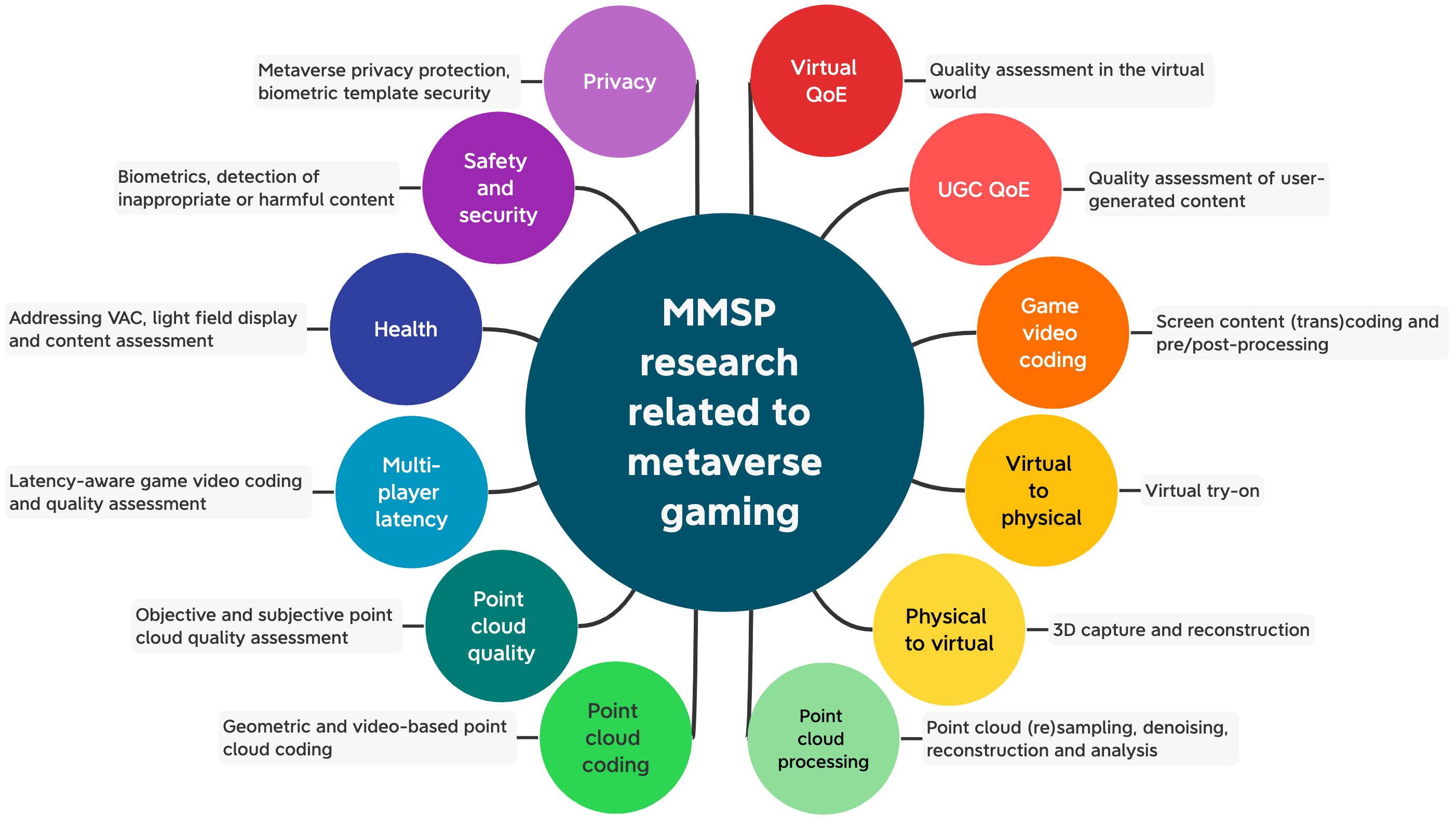}
\caption{MMSP research related to metaverse gaming.} 
\label{fig:summary_diagram}
\end{figure*}

\begin{table}[t]
    \centering
    \caption{Selected challenges related to metaverse gaming and corresponding MMSP research. Note that the list of challenges, research topics, and especially the list of references are far from exhaustive.}
    \label{tab:summary_of_challenges}
    \begin{tabularx}{0.48\textwidth}{>{\raggedright\arraybackslash}m{1.8cm} | >{\raggedright\arraybackslash}m{4.2cm} |>{\raggedright\arraybackslash}m{1.2cm} }
    \toprule
       \textbf{Challenge}  & \textbf{MMSP research} & \textbf{References} \\ \midrule
       Virtual QoE & Quality assessment in the virtual world & \cite{HEVC_VR_videos,foveated_compression_VR} \\ \hline
       UGC QoE & Quality assessment of UGC video & \cite{gaming_UGC_WACV,quality_gaming_TIP} \\ \hline
       Game video coding & Screen content coding,  transcoding, and pre/post-processing & \cite{HEVC-SCC,peng2016overview,bross2021overview,xu2021overview, SR_game_video_ICASSP2023,multitask_SCC_ISCAS2023} \\ \hline
       Virtual $\to$ physical & Virtual try-on & \cite{VITION_CVPR2018,OccluMix_TMM2023,AI_Fashion_SPM2023} \\ \hline
       Physical $\to$ virtual & 3D capture and reconstruction & \cite{RGBD_CVPR2015,RGBD_ICRA2019,Hartley2004} \\ \hline
       Point cloud processing & Point cloud (re)sampling, denoising,  reconstruction, and analysis & \cite{PC_sampling_TPAMI2023,PC_resampling_MMSP2021,PC_denoising_TIP2020,PC_registration_TMM2023,PCSR_TIP2022,3Dseg_ICME2017,GeoClass_TMM2022} \\ \hline
       Point cloud coding & Geometric and video-based point cloud compression & \cite{MPEG-PCC_SPM2019,MPEG-PCC_JETCAS2019,PCGC_MMSP2022} \\ \hline
       Point cloud quality & Objective and subjective quality assessment of point clouds & \cite{PCQE_QoMEX2019,PointXR_QoMEX2020,PCQA_MMSP2021,PCQM_QoMEX2020,PCQA_MMSP2022} \\ \hline
       Multiplayer latency  & Latency-aware game video coding and quality assessment & \cite{Foveated_game_MMSP2017,Game_video_MMSP2021,QoE_game_MMSP2014,GAMIVAL_SPL2023} \\ \hline
       Health   & Addressing VAC, light field display and content assessment & \cite{VAC_QoMEX2011,VAC_TIP2016,multifocal_ICME2016,multifocal_ICMEW2020,CTDM_TIP2023,LF_perceptual_QoMEX2017,LFQA_TIP2020} \\ \hline
       Safety and security  & Biometrics, detection of inappropriate or harmful content & \cite{face_rec_MMSP2007,head_mouth_MMSP2006,gait_MMSP2004,AV_MMSP1998,AV_MMSP2009,ApparelNet_MMSP2021,adult_ICME2009,adult_ICME2011,harmful_MMSP2014,adult_neurocomputing2017} \\ \hline
       Privacy & Player privacy, biometric template protection & \cite{MetaGuard,biotemplate_IEEEMM2014,biometric_protection_SPM2015,cancelable_biometrics_SPM2015} \\
    \bottomrule
    \end{tabularx}
\end{table}

\subsection{Privacy}

Privacy issues refer to unintended disclosure of private information that may be used to identify a person or some of their attributes (age, gender, etc.), which would enable someone to take advantage of them, impersonate them, or steal their identity. It is encouraging to see that metaverse privacy is on the research agenda and that solutions such as
MetaGuard~\cite{MetaGuard} are beginning to appear. MetaGuard employs differential privacy to make it more difficult for others to infer player's attributes such as age, gender, height, etc.

Since biometrics are an increasing part of metaverse safety and security measures, privacy issues related to biometric data are also important. In this context, it is worth mentioning the work on biometric template protection~\cite{biotemplate_IEEEMM2014,biometric_protection_SPM2015} and cancelable biometrics~\cite{cancelable_biometrics_SPM2015}, where the MMSP community has made important contributions.

\section{Summary and conclusions}
\label{sec:conclusions}

This paper has provided a young person's view of metaverse gaming. Related MMSP research challenges have been summarized in Table~\ref{tab:summary_of_challenges} and Fig.~\ref{fig:summary_diagram}. It is evident that MMSP research has an important role to play in the future development of the Metaverse.

\bibliographystyle{IEEEtran}
\bibliography{refs}

\end{document}